\documentstyle[12pt,epsfig]{article}

\newcommand{\beq}{\begin{equation}}
\newcommand{\eeq}{\end{equation}}
\def\beqa{\begin{eqnarray}}
\def\eeqa{\end{eqnarray}}

\def\nn{\nonumber}
\def\q{\quad}
\def\t{\tilde}
\def\th{\theta}
\def\mn{{\mu \nu}}
\def\bc{\begin{center}}
\def\ec{\end{center}}

\begin{document}

\begin{titlepage}
\begin{center}
\vspace{1cm}

\large {\bf Gravity in higher codimension de Sitter Brane Worlds}\\
\vspace*{5mm}
\normalsize

{\bf  I. Olasagasti and K.Tamvakis}

\smallskip
\medskip
{\it Physics Department, University of Ioannina,\\
GR-451 10 Ioannina, Greece}

\smallskip
\end{center}
\vskip0.6in

\centerline{\large\bf Abstract}

We study solutions of Einstein's equations corresponding
codimension $n>2$
global topological defects with de Sitter slices.
We analyze a class of solutions that are
cylindrically symmetric and admit
positive, negative or zero {\textit{Bulk}} cosmological constant.
We derive the relevant graviton equations.
For an extended brane, the properties of the solution depend on
appropriate boundary conditions that the exterior solutions have to
satisfy near the core. As an alternative we consider matching copies of
the exterior solution related by symmetry. We show that we can get
localization only when the {\textit{Bulk}} cosmological constant is negative. We
obtain a condition on the global defect symmetry breaking scale which
ultimately controls the size of the $n-1$ internal dimensions at the
position of the brane. The induced metric on the brane, in the case of
mirror spacetimes, is a direct product of a de Sitter space and an
$(n-1)$-sphere, while the metric of the embedding spacetime is a warped
product and the actual size of the $(n-1)$-sphere changes as we move
along the radial direction. The solutions possess naked singularities,
which nevertheless satisfy no-flow conditions.

\end{titlepage}

\section{Introduction}
The idea that the ordinary spacetime can be associated with a
{\textit{Brane}} embedded in a higher dimensional space has received considerable attention in the last few
years\cite{I}\cite{NIMA1}\cite{NIMA2}\cite{LISA1}.
This alternative to Kaluza-Klein compactification assumes the
localization of matter degrees of freedom on a topological defect or
{\textit{Brane}}\cite{SASHA}. Such assumptions are backed by expectations of matter
field localization through specific String Theory dynamics for
$D$-Branes\cite{EDWARD}. In contrast, gravity propagates in the {\textit{Bulk}} but
it retains its four-dimensional character through an effective
localization on the Brane resulting from the curvature of extra
dimensions\cite{LISA2}\cite{WHO}. Such is the case of a $3$-Brane in a $5$-dimensional $AdS$
spacetime where gravitation on the Brane is Newtonian with corrections
small at macroscopic scales. Models with more than one extra dimensions
have also been constructed based on higher {\textit{codimension or
transverse dimension}} defects. Although a number of such solutions are
known, arising either from local or global defects, graviton
perturbations have only been studied,
to the best of our knowledge,
in the case of flat
$3$-Branes\cite{Randjbar-Daemi:2002pq}, \cite{Csaki:2000fc},
\cite{Gregory:1995qh}, \cite{Charmousis:2001hg}, \cite{Benson:2001ac}.

There is a considerable interest in de Sitter space lately, to a large
extent motivated by the astrophysical evidence in favor of
a present phase of accelerated expansion.
Independently of that, there is plenty of
motivation to study de-Sitter Branes made up of scalar fields in the
{\textit{Bulk}} that form a global defect. In this paper we explore a class of
codimension $n>2$
spacetimes with non-flat $3$-Branes and
examine their gravitational excitation spectrum and its localization
properties.

Let us consider a $d$-dimensional spacetime composed of a
$q$-dimensional Brane embedded in $n$ extra dimensions ($d=n+q$). The
Brane corresponds to a global defect arising from $n$ scalar fields
$\phi^a$ interacting through a potential
$$V(\phi)=\frac{\lambda}{4}(\phi^a\phi^a-\eta^2)^2$$ Such a potential
allows for solutions with non-trivial mappings of the vacuum manifold,
the $(n-1)$-sphere $\phi^a\phi^a=\eta^2$, onto the $n$-dimensional space
transverse to the Brane. The simplest non-trivial configuration is
$$\phi^a=\eta f(r)\frac{y^a}{y}$$
with $y^2\equiv y^a y^a$,
$y^a=\{y \cos \th_1,y \sin \th_1 \cos \th_2,...   \}$,
and $\{\th_1,\dots,\th_{n-1}\}$ angular coordinates on the unit $(n-1)$-sphere.
$r$ is a radial coordinate on the space transverse to the defect
and for solutions representing a
defect the profile of $f(r)$ goes from a vanishing value at the center
of the {\textit{core}}
to $1$ at the outside. Defined in this
way the core of the defect is the region where the potential is
significantly different from zero. This is the region that defines the
Brane.

One can quite easily find solutions that solve Einstein's field
equations in the region outside the core/Brane, that is, where $f(r)=1$.
A complete spacetime solution should also include the solution for the
core region, which should me matched to the exterior solutions at some
matching surface. We consider the outside
region and see up to what point this might be sufficient in discussing
the localization of gravity. In addition to the
energy-momentum tensor for the scalar fields, we include
the effect of a {\textit{Bulk}}
cosmological constant $\Lambda$ whose sign we leave arbitrary.
For a class of known cylindrically symmetric solutions for the exterior
region we write the graviton equation and find the exact solutions.
Ultimately, in order to determine whether gravity is actually localized we
need to know the specific form of the core matter content and geometry.
Instead of fixing the matter on the core, we  proceed
by considering the
brane as an infinitely thin wall with $n-1$ internal
dimensions of finite size. We do this by matching two mirror copies
of the exterior solutions. In this picture although standard matter
is somehow localized along the transverse direction it can still
see the $n-1$ internal
dimensions.
We thus have two requirements in order
to get localized gravity. On the one hand we need to look at
the behavior in the direction transverse to the wall,
and on the other, just as in standard Kaluza-Klein dimensional reduction,
we need the internal dimensions at the wall position to be small enough
so that modes wrapping around them are correspondingly massive.
This thin wall approach allows us to study the spectrum in more detail,
even if we keep our analysis qualitative.
For the class of solutions at hand, we find  that
gravity can be localized only for a negative {\textit{Bulk}} cosmological constant
and for a certain range of the global defect parameters.

Here is the plan of the paper. In section 2 we present the class of cylindrically
symmetric solutions, taken
from \cite{Olasagasti:2000gx} with appropriate reparametrizations. They
correspond to solutions with de Sitter slices and backgrounds with negative, positive
and zero {\textit{Bulk}} cosmological constant.
In section 3 we write the relevant equations for
the graviton in warped geometries with a spherically symmetric section.
In section 4 we give the radial equation and solve it. We also
write the boundary conditions set by
the specific nature of the matter making up the core.
In section 5 we construct and analyze the mirror spacetimes mentioned above. We explain our conclusions
in section 6.

 \section{Cylindrically symmetric Branes}
We begin by writing down the solutions that we are going to take as
background solutions. They are represented by
cylindrically symmetric metrics of the form
 \beq
ds^2= dr^2+e^{2A(r)}\left(L^2 d\Omega_{n-1}^2+\hat{g}_{\mu\nu}dx^{\mu}dx^{\nu}\right)
 \eeq
where $r$ is the radial coordinate in the space transverse to the Brane
and $d\Omega_{n-1}^2$ is the line-element for a unit radius
$(n-1)-$dimensional sphere. We shall take the metric $\hat{g}_{\mu\nu}$
on the Brane to be that of an $q-$dimensional de Sitter space. The
curvature of this de Sitter space will be parametrized in the usual way
as $\hat{R}_{\mu\nu}=(q-1)H^2\hat{g}_{\mu\nu}$ and $\hat{R}=q(q-1)H^2 $.
The associated energy-momentum tensor is

 $$T_r^r=-\left(\frac{(n-1)\eta^2}{2 L^2}\;e^{-2A}+\Lambda\right)
 \,\,,\,\,\,\,\,T_{\mu}^{\nu}=\delta_{\mu}^{\nu}T_r^r\,\,,\,\,\,\,\,
 T_j^i=-\delta_j^i\left(\frac{(n-3)\eta^2}{2 L^2}\;e^{-2A}+\Lambda\right)$$

The following solutions exist \cite{Olasagasti:2000gx}

\beq
e^{2A(r)}=\left\{\begin{array}{cc}
\alpha_{+}^2\sin^2\left(\beta_+(r_0-r)\right)\,\,&\,\,\,(\Lambda>0)\\
\alpha_0^2(r-r_0)^2\,\,&\,\,\,(\Lambda=0)\\
\alpha_-^2\sinh^2\left(\beta_-(r_0-r)\right)\,\,&\,\,\,(\Lambda<0)
\end{array}\right.
\q\q  L^2= {(n-2-\kappa^2 \eta^2) \over (q-1) H^2}
\label{sols}
\eeq
For the first and the third cases we have
\beq
\alpha_{\pm}^2=\frac{(q-1)(d-1)H^2}
{2\kappa^2|\Lambda|}
\,\,\,,\,\,\,\,\,\,\,
\beta_{\pm}^2=
\frac{2\kappa^2|\Lambda|}{(d-2)(d-1)}
\eeq
while in the second case
\beq
\alpha_0^2=\frac{(q-1)}{(d-2)}H^2
\eeq
$\kappa^2\equiv 8\pi G$  gives the gravitational coupling on the {\textit{Bulk}}.

We note that to be well defined we need $n\geq 3$.
At some value of $r$
the above exterior solutions should match the undetermined core solution.
Without loss of generality we will take this matching hypersurface
to be at $r=0$.
We also note that these geometries
are characterized by a naked singularity at $r=r_0$. At that point
the energy-momentum tensor diverges. This should not come as a
surprise since this corresponds to the vanishing of the radius of
the $(n-1)$-dimensional spheres.

In the above $H$ is an arbitrary parameter. The effective $q-$dimensional curvature
changes along $r$ as
\beq
\hat R_{ef}=\hat R e^{-2 A}=
{2 q \kappa^2 \Lambda\over (d-1)\sin^2(\beta_+ (r_0-r))},\q
{q (d-2)\over (r_0-r)^2},\q
{2 q \kappa^2|\Lambda|\over (d-1)\sinh^2(\beta_- (r_0-r))}
\eeq
for $\Lambda>0,\;\Lambda=0,\;\Lambda<0$,
which clearly does not depend on $H$. However we find it convenient
to keep it. We can fix it by demanding that it gives the
actual Hubble parameter at a given radial position.
For example,
by demanding that it gives the
actual Hubble parameter at $r=0$ we have
\beq
H^2=
{2 \kappa^2 \Lambda\over (q-1)(d-1)\sin^2(\beta_+ r_0)},\;
{ (d-2)\over  (q-1) r_0^2},\;
{2  \kappa^2 |\Lambda|\over  (q-1) (d-1)\sinh^2(\beta_- r_0)}
\label{hubble}
\eeq
This is clearly equivalent to normalizing the
warp factor so that we have $e^{A}=1$ at $r=0$.

\section{Graviton equation}

When considering gravitational perturbations for a gravity-scalars
system like the global defect we
have considered, the scalar variations $\delta\phi^a$,
for a transverse, traceless graviton in the harmonic gauge,
decouple\cite{DeWolfe:1999cp} from the graviton and, thus, can be ignored.
In the present section we will derive the equation that corresponds
to this traceless graviton for a
general warped geometry of the type
\beq
ds^2=e^{2A(y)} d\hat{s}^2+d\tilde{s}^2\equiv
e^{2A(y)} \hat{g}_\mn dx^{\mu} dx^{\nu}+\gamma_{ab} dy^a dy^b
\label{warp}
 \eeq
where $d\hat{s}^2$ is $q-$dimensional
and $d\tilde{s}^2$ $n-$dimensional.
For such metrics we can write the perturbation modes as a product
$$h_\mn(x,y)\equiv e^{2A(y)}\phi(y)\epsilon_\mn(x)$$
and one finds that the equation for $\phi$ coincides
with the equation for a scalar field, namely,
\beq
\t\nabla^2 \phi+q \t\nabla A \; \t\nabla \phi+e^{-2A} \;m^2\phi
=0.
\label{trans}
\eeq
where $m^2$ is the mass term of the $q-$dimensional spin-two perturbations and
the tildes denote geometrical quantities calculated
using the metric $d\tilde{s}^2$ of the extra dimensions
\footnote{The {\textit{ordinary-space}} graviton factor
 satisfies the equation \cite{Garriga:1999yh}
 $$-\frac{1}{2}\hat{\nabla}^2\epsilon_{\mu\nu}
 +\left(-H^2+\frac{m^2}{2}\right)\epsilon_{\mu\nu}=0$$}.

Since the solutions that we want to study are cylindrically symmetric,
we should specialize to the class of metrics that can
be written as (\ref{warp}) with

\beq
d\t s^2=\gamma_{ab} dy^a dy^b=dr^2+R^2(r) d\Omega^2_{n-1}
\label{warp2}
\eeq
where $r$ is a radial coordinate in the space transverse to the brane and
$d\Omega^2_{n-1}$ defined as in the previous section.
Because of
the symmetries of the solution it is natural to write
$\phi(y)=\psi(r) Z_{\ell}(\Omega)$
where $Z_{\ell}(\Omega)$ are spherical harmonics in
$S^{(n-1)}$. The equation they satisfy is
\beq
{\Box}Z_{\ell}+\ell(\ell+n-2)Z_{\ell}=0
\eeq
where $\Box$ here is the box operator on $S^{(n-1)}$. Plugging this into (\ref{trans})
we get the following equation for
$\psi(r)$

\beq
\psi'' +\psi'\left( q\;{ A'}
+(n-1){ R' \over R}\right)
+\left(m^2e^{-2A}-{\ell(\ell+n-2)\over R^2}\right)\; \psi=0
\eeq

For the particular solutions displayed in the previous section we had
$R(r)=L e^{A(r)}$.
In this case
 the equation simplifies even further reducing to
 \beq
 \psi^{''}+(d-1)A'\psi'+\left(m^2-{\ell(\ell+n-2)\over L^2}\right)e^{-2A}\psi=0
 \eeq

 The normalization of the modes is given by
the term  \cite{Randjbar-Daemi:2002pq}
\[
\int e^{(d-3)A}\psi^2 \; dr \;
\int L^{(n-1)} Z^2_{\ell}(\Omega)\;d\Omega_{n-1}
\int \sqrt{\hat g} \hat g^\mn \epsilon_{\lambda \kappa,\mu}
\epsilon_{\lambda \kappa,\nu}
\]
So, the condition for normalizable $q-$dimensional modes, in our choice of metric, reduces to
\beq
\int dr\,e^{(d-3)A(r)}\psi^2(r)<\infty
\eeq
The normalization integral includes the core region
where we do not know the solution, and extends up to the limiting singular point $r_0$
of the outside region.

The zero modes ($m^2=\ell=0$) for the solutions displayed in the
previous section can be easily derived from the above equation.
Imposing the normalization condition for them we see that the integral
between $r=0$ and $r=r_0$ is finite only for the constant
zero mode solution $\psi_0(r)=const.$. The other linearly independent
choice gives a divergent contribution to the normalization integral due to its
behavior near the singularity at $r=r_0$.

In order to study the massive spectrum it is convenient to
transform the graviton equation into a Schrodinger type differential equation.
This can be achieved by introducing  a new variable $z$ and a new
graviton function $\chi(z)$ defined through
\beq
dz=e^{- A}dr\,\,\,,\,\,\,\,\chi=e^{(d/2-1)A}\psi
\label{conf}
\eeq
In the following we will refer to this choice as the conformal gauge.
After these transformations
the graviton equation takes the Schrodinger form
\beq
\left\{-\frac{1}{2}\frac{d^2}{dz^2}+V(z)\right\}\chi(z)=\frac{m^2}{2}\chi(z)
\eeq
where the {\textit{potential}} $V$ is defined in terms of the warp function as
\beq
V=-\frac{1}{4}(d-2)\left(\ddot{A}-\frac{1}{2}(d-2)\dot{A}^2\right)
+\frac{1}{2L^2}\;\ell(\ell+n-2)
\eeq
Moreover, for $\chi(z)$ the normalizability condition is the usual one, that is,
\beq
\int\chi^2(z) \, dz<\infty.
\eeq
\section{Spectrum}
We can work the transformation to the conformal gauge
for each of the solutions
presented in the previous section. In the case of positive {\textit{Bulk}} cosmological
constant $\Lambda>0$ we obtain
\beq
e^{2A}=\frac{\alpha_+^2}{\cosh^2(\t\alpha z)}
\eeq
in terms of $\tilde{\alpha}^2\equiv \alpha_+^2\beta_+^2=(q-1)H^2/(d-2)$.
The variable $z$ takes values in the range $z_0<z<\infty$, where $z_0\equiv z(0)$.
At the singularity $r=r_0$,
the variable $z$ reaches infinity. The potential as a function of $z$ is
\beq
V(z)=\frac{1}{2L^2}\;\ell(\ell+n-2)
+\frac{\tilde{\alpha}^2}{8}
\left\{(d-2)^2-\frac{d(d-2)}{\cosh^2(\tilde{\alpha}z)}\right\}
\eeq

The general solution of the above Schrodinger problem is given
in terms of hypergeometric functions as
\beq
\chi^{(\mp)}(z)=\cosh(\t \alpha z)^{(\mp \sqrt{c})} \;
{}_2F_1[{d/ 4}\pm {\sqrt{c}/2},{(2-d)/ 4}\pm {\sqrt{c}/2}\;
;1\pm \sqrt{c}\; ;{1/ \cosh^2(\t \alpha z)}]
\eeq
with $c$ used as a shorthand for
\beq
c={(d-2)^2 \over 4}+
{(d-2)\ell(\ell+n-2)\over (n-2-\kappa^2\eta^2)}-{(d-2)\;m^2\over (q-1) H^2}
\label{c}
\eeq
To check normalizability we look at the behavior as $z\rightarrow \infty$
\beq
\chi^{(\pm)}(z)\sim e^{\pm \sqrt{c}\t\alpha z}
\eeq
The behavior changes qualitatively at $m=m_c$, which is the value that makes
$c=0$
\beq
m^2_c=H^2\;{(q-1)\over 4}\left\{ (d-2)+{4\ell(\ell+n-2)\over n-2-\kappa^2\eta^2}\right\}
\label{mc}
\eeq
For $c<0$, or equivalently $m^2>m_c^2$, asymptotically
the functions approach  plane waves
\beq
\chi^{(\pm)}(z)\sim e^{\pm i\sqrt{|c|}\tilde{\alpha}z}
\eeq
 so both will in principle lead to
acceptable continuum  modes. For $c>0$, or
$0<m^2<m^2_c$, only one type of modes is
normalizable, namely,
\beq
\chi^{(-)}(z)\sim e^{-\sqrt{c}\tilde{\alpha} z}.
\eeq
These modes correspond to the discrete part of the massive spectrum.
The number of discrete modes and their mass eigenvalues will
depend on $z_0$ and the boundary conditions at that point.
Note that for $m^2=0, \ell=0$, which corresponds to $c=(d-2)^2/4$, we recover the
$\psi_0(r)=const.$ zero-mode.

For a vanishing {\textit{Bulk}} cosmological constant $\Lambda=0$ the variables are related as
\beq
r_0-r=r_0e^{-(z-z_0)\alpha_0}
\eeq
 while the range of $z$
is as in the previous case. Here we arrive at a constant potential
\beq
V=\frac{\alpha_0^2}{8}\;(d-2)^2+\frac{1}{2L^2}\;\ell(\ell+n-2)
\eeq
Again, the solutions depend on the parameter $c$ defined above
in (\ref{c}). For $c>0$
we have a continuum of plane wave solutions
\beq
\chi^{(\pm)}_m(z)=N_me^{\pm i\alpha_0\sqrt{|c|}z}
\eeq
 defined in the half line $z_0<z<\infty$. The bounded solutions arising for $c>0$
 \beq
 \chi^{(-)}(z)=N^{(-)}e^{-\alpha_0\sqrt{c}z}
 \eeq
 are subject to the appropriate boundary conditions at $z_0$.

 Similarly, in the case of negative {\textit{Bulk}} cosmological constant $\Lambda<0$, we have
 \beq
 e^{2A}=\frac{\alpha_-^2 }{\sinh^2(\tilde{\alpha}z)}
 \eeq
 for the same range of parameter $z_0<z<\infty$  and using again
 $\tilde{\alpha}^2=\beta_-^2\alpha_-^2=(q-1)H^2/(d-2)$.
It is straightforward to obtain the potential as
 \beq
V(z)=\frac{1}{2L^2}\;\ell(\ell+n-2)
+\frac{\tilde{\alpha}^2}{8}\left\{
(d-2)^2+\frac{d(d-2)}{\sinh^2(\tilde{\alpha}z)}
\right\}
\eeq
 It is clear that for $m^2>m_c^2$ (see eq (\ref{mc})),
 we obtain again the continuum of asymptotic plane waves. The existence of
 acceptable normalizable modes in the interval $[0,m_c^2]$
 depends on the boundary conditions to be imposed at $z_0$.
 The exact form of the two
 linearly independent solutions is given by
\beq
\chi(z)^{(\mp)}=\sinh(\t \alpha z)^{d/2} \cosh(\t \alpha z)^{-(d\pm2 \sqrt{c})/2}
{}_2F_1[{d/ 4}\pm {\sqrt{c}/2},{(d+2)/ 4}\pm {\sqrt{c}/2}\;
;1\pm \sqrt{c}\; ;{1/ \cosh^2(\t \alpha z)}]
\eeq
with $c$ defined as before.

The discrete part of the spectrum, including the zero mode,
in all the above three cases rests on the boundary conditions at $z_0$, where
the outer solutions should match the core solutions.
One expects that a regular solution at the center of the core will be possible
only for a discrete set of values of $m^2$.

As $z\rightarrow \infty$ we encounter the naked singularity. For acceptable solutions
we impose unitary boundary conditions, which amount to a no flow condition
into the singularity for conserved quantities. From the invariance due to the
symmetries along the brane one obtains the
constraint
\beq
\mathop{\lim}\limits_{z \to \infty}
 e^{(d-2)A(z)}\psi(z)\psi'(z)=0
\eeq

We have $\psi(z)=e^{-(d-2)A(z)/2}\chi(z)$ and as $z\rightarrow \infty$
for all three cases the
discrete modes go as
$\psi(z)\propto e^{-(d-2)\alpha z/2-\alpha\sqrt{c}z}$ so that the above
turns into
\beq
\mathop{\lim}\limits_{z \to \infty}
 e^{-\alpha\sqrt{c}z}=0,
\eeq
so the condition is fulfilled. For the continuum modes on the other hand ($c<0$)
$\psi(z)\propto e^{-(d-2)\alpha z/2-i\alpha\sqrt{|c|}z}$ so now the left hand side
of the condition reads
\beq
\mathop{\lim}\limits_{z \to \infty}
 e^{-i\alpha\sqrt{|c|}z}.
\eeq
This is a wildly oscillating function and corresponds to a function $\chi$ which
is not normalizable in a strict sense. As usual realistic states will be described
by a superposition that leads to a normalized function which means that
$\chi\rightarrow 0$ as $z\rightarrow \infty$
and therefore the condition will be fulfilled as well.
Normalizable modes therefore automatically satisfy the no-flow condition
and we do not have to worry about new boundary conditions imposed by
the singularity.

One can be more specific about the boundary conditions set by the interior
solution on the metric functions outside \cite{Gherghetta:2000jf}.
Taking into account the symmetry
of the solution we can expect the $T_{AB}$ components through the interior
to be
\beq
T^r_r=f_r(r), \q\q T^\mu_\nu=\delta^\mu_\nu f_0(r),
\q\q T^\theta_{\theta'}=\delta^\theta_{\theta'} f_\theta(r)
\eeq

The exterior solution extends from $r=0$, the matching hypersurface,
to the singularity at $r=r_0$. The core region, on the other hand
extends from $r=-r_c$ which gives the center of the solution,
to the hypersurface $r=0$.
If we integrate Einstein's equations,
in the form $R^a_b=\kappa^2[ T^a_b-\delta^a_b T/(d-2)]$,
through the core region $-r_c<r<0$
we obtain
\beqa
R'R^{n-2}e^{qA}|^0_{-r_c}&=&{\kappa^2\over d-2}[\mu_r+q\mu_0-(q-1)\mu_\theta]+
(n-2)\int^0_{-r_c} R^{n-3}e^{qA} d\rho \label{bc1}\\
A'e^{qA}R^{n-1}|^0_{-r_c}&=&{\kappa^2\over d-2}[\mu_r-(n-2)\mu_0+(n-1)\mu_\theta]
\nn \\
&&+(n-2-\kappa^2 \eta^2)\int^0_{-r_c} R^{ n-1}e^{(q-2)A} d\rho
\label{bc2}
\eeqa
where
\beq
\mu_i\equiv \int^0_{-r_c} R^{n-1}e^{q A} f_i(\rho) d\rho
\eeq
and $i=(r,0,\theta)$ 

To have a regular solution at the center, $r= -r_c$,
we will have the following conditions
\beq
R'(-r_c)=1, \q  A'(-r_c)=A(-r_c)=0
\eeq
These can be used in (\ref{bc1}),(\ref{bc2}) so that
the left hand side will give the
appropriate boundary conditions at $r=0$.

Above we have given exact solutions for the modes in the outside region.
To determine the shape of gravity in the brane it is important
to know the number and masses of the discrete modes and this in turn
requires knowledge about the detailed nature of the core. Instead of
choosing a possible core matter we prefer to  study an alternative set-up,
where we consider an infinitely thin wall at $z=0$ matching
two mirror copies of the same spacetime. This is what we will call
the mirror spacetimes.

\section{Mirror spacetimes}
Another way to model branes with higher codimension is to consider an infinitely thin wall
just as in the codimension one case. The difference being that the wall will have
the topology ${\cal M}_q\times S^{n-1}$. The infinitely thin wall is thus a factorized space
unlike the embedding spacetime which is not generally factorizable this way.
In the present section we shall consider matching two copies of the above manifolds
across the ${\cal M}_q\times S^{n-1}$ wall (for simplicity,
we are  assuming $Z_2$ symmetry across the wall).

It is well known that for an infinitely  thin wall the metric is
continuous across the wall but that the extrinsic curvature is
discontinuous. We can characterize this discontinuity by the jump
of the extrinsic curvature across the wall, which is related to
the localized energy momentum tensor through
\beq
[K]^i_j-\delta^i_j Tr[K] =-\kappa^2 S^i_j
\eeq
where $S_{ij}$ is
the localized energy momentum tensor, $K_{ij}$ the extrinsic
curvature  and the brackets denote the difference of a quantity at
the two sides of the wall.

We can consider the present set-up as obtained from the one in the previous
sections by
substituting what we referred to as the interior region by a mirror copy
of the exterior solution. Consequently the former smooth matching surface turns into
a brane with an induced localized energy-momentum tensor.

For the metrics at hand (\ref{warp}),(\ref{warp2})
but in the conformal gauge eq(\ref{conf}), the explicit expression
for the extrinsic curvature reads
\beq
K^\mu_\nu=\delta^\mu_\nu {A'_w}e^{-A_w} \equiv \delta^\mu_\nu K_0, \q \q
K^\alpha_\beta=\delta^\alpha_\beta {R'_w\over R}e^{-A_w}
 \equiv \delta^\alpha_\beta K_\theta.
 \label{jump}
\eeq
where primes denote derivatives with respect to the conformal radial
coordinate $z$.
We will assume that
\beq
S^\mu_\nu=\delta^\mu_\nu S_0\q \q
S^\alpha_\beta=\delta^\alpha_\beta S_\theta
\eeq
which is consistent with the symmetry of our metric {\em ansatz}. From this we have that
\beqa
\kappa^2 S_0 & =&[(q-1)K_0+(n-1)K_\theta]\\
\kappa^2 S_\theta&=&[q K_0+(n-2)K_\theta]
\eeqa

A vacuum wall of tension $\sigma$ is characterized by a localized
energy-momentum tensor that is proportional to $h_{ij}$,
the induced metric on the wall, so that
$S_{ij}=-\sigma h_{ij}$.
We can see from the above that this relation can only be satisfied
if $K_0=K_\theta$.
Such vacuum walls will not be obtained
in general warped spacetimes with different warp factors
for the $q$-dimensional and $(n-1)$-sphere parts.
In that case the wall will have a non-uniform tension, different for
$x^\mu$ and the angular directions. This is an interesting case to investigate
since, unlike the solutions that we are treating in this paper, they
are not conformal to factorized spacetimes. We expect to investigate
this issue in the future.

Let us stay with the conformally factorized spacetimes.
If we consider the solutions used earlier in the paper,
we will indeed obtain a vacuum wall when we match two copies of the
spacetime.
Since
in order to represent a positive tension wall we will need
the warp factors to decrease away from the wall, we can
paste two mirror copies of the solutions for
$z$ in the range $z_0<z<\infty$ matching along the
hypersurface given by $z=z_0$ ($r=0$ in the original coordinates).

The tension $\sigma$  of this wall will be given by
\beq
\kappa^2 \;\sigma=-(d-2)e^{-A(z_0)} A'(z_0)
\eeq

As was the case
above, we will encounter naked singularities at a finite distance from
the brane.

When we go on to study the
graviton perturbations, the effective potential
at either side of the wall will be mirror copies of the potentials obtained
above and in addition we will have a term
proportional to a $\delta$ function,
which is the fingerprint of the discontinuity introduced by the wall
\beq
-\chi''+
\left[V_{r}+\left({(q-1)\over 2} {[A']}+{(n-1)\over 2} {[R']\over R_0}\right)
\delta(z-z_0)
\right] \chi
=m^2 \chi
\eeq
where $V_r$ is the regular part of the potential, which has been calculated above
for each of the spacetimes.

\begin{figure}[h]
\epsfbox[90 68 520 170]{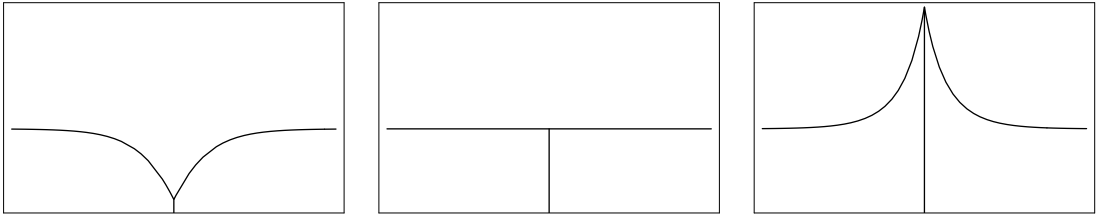}
\caption{Thin wall potentials for $\Lambda>0$, $\Lambda=0$ and $\Lambda<0$ respectively. }
\label{fig1}
\end{figure}

The delta function gives a boundary condition at the position of the wall.
In order to satisfy the above equation,
 $\chi$ must be continuous but its derivative must have
a discontinuity. The jump of the derivative across the wall is given by
\beq
{[\chi'] \over \chi_0}={(q-1)\over 2} {[A']}+{(n-1)\over 2} {[R']\over R_0}
={(d-2)\over 2} {[A']}
\label{bc3}
\eeq
where the last equality is valid only for the present conformally factorizable
spacetimes. The boundary condition is also satisfied
by modes with $\chi_0=0,\;[\chi']=0$, however these are no concern to
us since they are zero at the position of the brane and thus
do not couple to matter sitting there.

Eq (\ref{bc3}) gives an explicit boundary condition at the wall. As we will show
below, it is trivially satisfied by the zero mode $\ell=0,m=0$ in all three
cases ($\Lambda>0$, $\Lambda=0$, $\Lambda<0$).
In Fig.(\ref{fig1}) we can see the form that the thin-wall potentials
take for each
sign of the {\textit{Bulk}} cosmological constant.

For $\Lambda>0$ the boundary condition (\ref{bc3}) is
\beq
{\chi'_0 \over \chi_0}=-{d-2\over 2}\; \t \alpha \tanh(\t \alpha z_0)
\label{bc3a}
\eeq

As we showed earlier in the paper the discrete and continuum parts of the spectrum
are separated by $c=0$. We showed that when $c>0$ only one of the modes
is normalizable, if we now impose the boundary condition set by the wall,
only some values of $c$ will be allowed, thus leading to the discrete spectrum.
As is apparent from Fig.(\ref{fig1}), we can expect a number of discrete modes
for different values of $c$.

\begin{figure}[h]
\epsfbox[90 68 520 152]{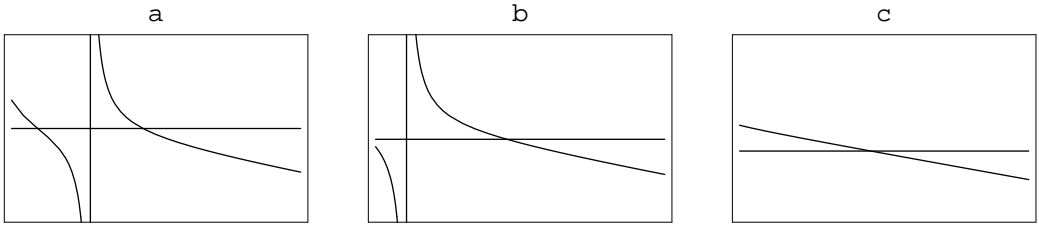}
\caption{boundary condition equation for $\Lambda>0$. The depth of the
well decreases from left to right }
\label{fig2}
\end{figure}

In Fig.(\ref{fig2}) we represent both sides of equation (\ref{bc3a})
as a function of $c$
for three extra dimensions ($d=7$) and three different values of the wall
tension. We can see that there is a critical value of the
tension $\sigma_+$ such that for $\sigma<\sigma_+$ there are two values
of $c_s$ (with $s$ labelling the values)
that satisfy the boundary condition. These lead to two towers
of states if we allow for $\ell \neq 0$ modes
\beq
m^2(\ell,s)=H^2\;{(q-1)\over 4}\left\{ (d-2)-{4c_s\over (d-2)}
+{4\ell(\ell+n-2)\over n-2-\kappa^2\eta^2}\right\}
\eeq
where $c_1=(d-2)^2/4$ corresponds to the lowest lying value which leads
to
\beq
m^2(\ell,s=1)=H^2\;{(q-1)\ell(\ell+n-2)\over n-2-\kappa^2\eta^2}
\label{zeromode}
\eeq
these modes include the zero mode, $\ell=0$, and then
modes with masses \mbox{$m^2\sim H^2/(n-2-\kappa^2\eta^2)$} which will
 become more and more massive as
$\kappa^2 \eta^2 \rightarrow n-2$.

For $\Lambda=0$ the boundary condition (\ref{bc3}) is
\beq
{\chi'_0 \over \chi_0}=-{d-2\over 2}\; \t \alpha
\eeq
and only the $c_1$ value satisfies the boundary condition for
$m^2<m_c^2$.

For $\Lambda<0$ the boundary condition (\ref{bc3}) is
\beq
{\chi'_0 \over \chi_0}=-{d-2\over 2}\; \t \alpha \coth(\t \alpha z_0)
\eeq
Again only the modes with $c=c_1$ satisfy the condition when $m^2<m_c^2$.

What do we get from the above?. As is well known, because we have a de Sitter brane
we have a mass gap for the continuum spectrum. The continuum
starts at $m=m_c$ which is of the order of $H$ for $\ell=0$.
We recall that $H$ is the actual Hubble parameter at the wall and
is given by eq (\ref{hubble}). From the $q$-dimensional
perspective the lightest of these modes
will be suppressed by their
mass only on superhorizon length scales $L> 1/m_c\sim 1/H$.
This means that unless their wave functions are suppressed on the brane
these modes will lead to higher dimensional gravity. As we see in Fig.(\ref{fig1})
there is no suppression for $\Lambda>0$ or $\Lambda=0$ so we can
conclude that gravity is not localized in these cases. As is the case
in the 5-dimensional thin wall case, it seems that only the
$\Lambda<0$ can do the job.

In this last case we should take care of the modes
trapped by the wall with masses given by eq (\ref{zeromode}). In addition to
the zero
mode with $\ell=0$, we have all the others with $\ell\neq0$. It is
thus important that these are massive enough, from the $q$-dimensional
point of view, so that at observable distances on the brane they
can be neglected. This can always be achieved by a symmetry breaking scale
$\eta$ sufficiently close to the critical value
$\kappa^2\eta^2_c\equiv (n-2)$.

We therefore conclude that in the mirror spacetimes gravity can only be localized
when the {\textit{Bulk}} cosmological constant is negative.
We expect the same to be true for the solutions with a realistic core as
well.

\section{Conclusions}
In the present paper we have extended previous analyses on the gravity
on braneworld models
by considering thick branes embedded in spacetimes with $n\geq 3$ extra dimensions
and, most importantly, by allowing a non-flat brane geometry. In particular
we have considered a brane geometry with $n-1$ {\em small} compactified dimensions
and q-dimensional de Sitter slices. We refer to these branes as thick since
they
do not correspond to a perfectly localized source in the extra dimensions.
In the first part we have modelled the branes as  a global
topological defect associating the brane with the core.
The sources in the {\textit{Bulk}} are a cosmological constant and the scalar fields
that make up the defect.
The defect solution is characterized by the fact that at some
distance from its center the potential energy reaches a minimum.
We refer to this region where the fields have their vacuum values as
the exterior region. We have
analyzed in some detail the solution in this exterior region
 with an eye at the localization of the graviton.
To this end we have written the relevant equations for
the metric perturbations around backgrounds
described by warped geometries.
 We have
further specialized the equations for  the case at hand,
that of spherically symmetric solutions
as seen from the extra dimensions. We have derived
the relevant equation for the radial dependence of
the perturbations
and then applied it to the solutions at hand.

We have assumed that the exterior solutions match at some
boundary hypersurface an unknown interior solution describing
the core of the defect, that is, what we consider to be
the brane. Focusing on the exterior part of the solution
we have found some necessary conditions for gravity to be
localized. They are not sufficient because
the specific nature of the core
will affect the modes through the boundary conditions
at the matching surface. The degree of arbitrariness associated with
 choosing a suitable interior solution describing the defect and matching it to
 the exterior solution can be circumvented by going to infinitely thin branes.
 Meaningful and general conclusions can still be drawn in this approach.
 Thus, in the last part of the paper
we have turned our attention to mirror spacetimes.
We have taken each of the exterior solutions and
we have constructed new solutions by pasting two mirror
copies of the same spacetime across a hypersurface, thus
creating a thin wall discontinuity which can be associated with the brane.
Since we
choose a hypersurface orthogonal to the radial direction
the resulting wall/brane has topology $M_{q} \times S^{n-1}$.

This is not equivalent to
considering the known 5D braneworlds with a number
of extra compactified dimensions. This would be the case
if the $d$-dimensional solution was a factorized geometry of the form
$M_{q+1} \times S^{n-1}$. Instead we have a warped geometry.
This means that the size of the $S^{n-1}$ part is not constant
throughout the spacetime. This new approach allowed us to solve the problem completely
since the presence of the infinitely thin wall fixes the
required boundary conditions.
For the three signs of the {\textit{Bulk}} cosmological constant,
the continuum spectrum starts at $m\sim H$. Since
$H$ gives the Hubble scale on the wall, these massive modes
won't be suppressed on the wall within the horizon and in order to
obtain q-dimensional gravity they should couple very weakly to the
brane which only happens for $\Lambda$  negative. This is not enough however. The
 brane in the mirror spacetimes
has $n-1$ internal dimensions and there are massive modes that wrap around these
whose function along the radial direction
is the same as the zero mode and consequently are not suppressed on the brane.
They will destroy localized gravity unless they are massive enough so that
at observable distances on the brane they can be integrated out.
For the solutions considered here the mass of these modes in units
of the  Hubble parameter mass scale are  set by
the symmetry breaking scale of the defect. It turns out that they can
be made arbitrarily massive by a symmetry breaking scale $\eta^2$
that approaches the critical value $(n-2)/\kappa^2$. For such
values of $\eta$, and always when $\Lambda<0$,
we conclude that gravity will be localized.

The solutions analyzed here possess
naked singularities. Nevertheless, we have argued that these
can be circumvented by imposing
no-flow boundary conditions. Indeed, we find that the no-flow condition
is automatically satisfied by normalizable modes\cite{KOT}. The solutions
studied here are conformally factorizable and, in that sense, quite simple.
We hope to study the gravity of more general solutions, which are not
conformally factorizable, in the near future.

{\textbf{Acknowledgements}}

The authors acknowledge support from the RTN programme HPRN-CT-2000-00152.
They also wish to thank the CERN Theory Division for its hospitality.


\end{document}